\documentclass[12pt,a4paper]{article}
\usepackage[utf8]{inputenc}
\usepackage[affil -it]{authblk}

\usepackage{graphicx}
\usepackage{amsmath,amsfonts,amssymb,textcomp}
\usepackage{color}
\usepackage[pdfencoding=auto, psdextra]{hyperref}
\hypersetup{colorlinks,linkcolor={blue},citecolor={blue},urlcolor={red}}  
\usepackage[version=4]{mhchem}

\title{RESONANCES IN THE SOLAR NEUTRINO CAPTURE CROSS-SECTION FOR $ \ce{^{76}Ge} $ NUCLEI }
\author[1,2]{A.\,N.\,Fazliakhmetov}
\author[1]{L.\,V.\,Inzhechik}
\author[1]{G.\,A.\, Koroteev}
\author[3]{ Yu.\,S.\, Lutostanky}
\author[3]{ V.\,N.\, Tikhonov}
\author[1,2]{ A.\,K.\, Vyborov}

\affil[1]{Moscow Institute of Physics and Technology, Dolgoprudny, Russia }
\affil[2]{Institute for Nuclear Research of Russian Academy of Sciences, Moscow, Russia}
\affil[3]{National Research Centre "Kurchatov Institute", Moscow, Russia}

\begin{document}
\maketitle

\begin{abstract}
The calculation results of solar neutrino capture cross-section on  the $ \ce{^{76}Ge} $ nucleus are presented. The charge-exchange strength function $S(E)$ was obtained from the experiment data in the $ \ce{^{76}Ge}( \ce{^{3}He}, t)\ce{^{76}As} $ reaction. The calculation includes the effect of the resonant structure of the strength function $S(E)$ on the calculated cross section $\sigma(E_{\nu})$. It is shown that only the giant Gamow-Teller resonance contributes about 20\%, and an even greater contribution is made by excited states located lower in the continuous part of the spectrum. These contributions should be taken into account in the calculation of background events in experiments on double beta decay of the GERDA type (for example LEGEND).
\end{abstract}

\section{Introduction}

\begin{figure}[h!]
\center
	\includegraphics[width=0.9\linewidth]{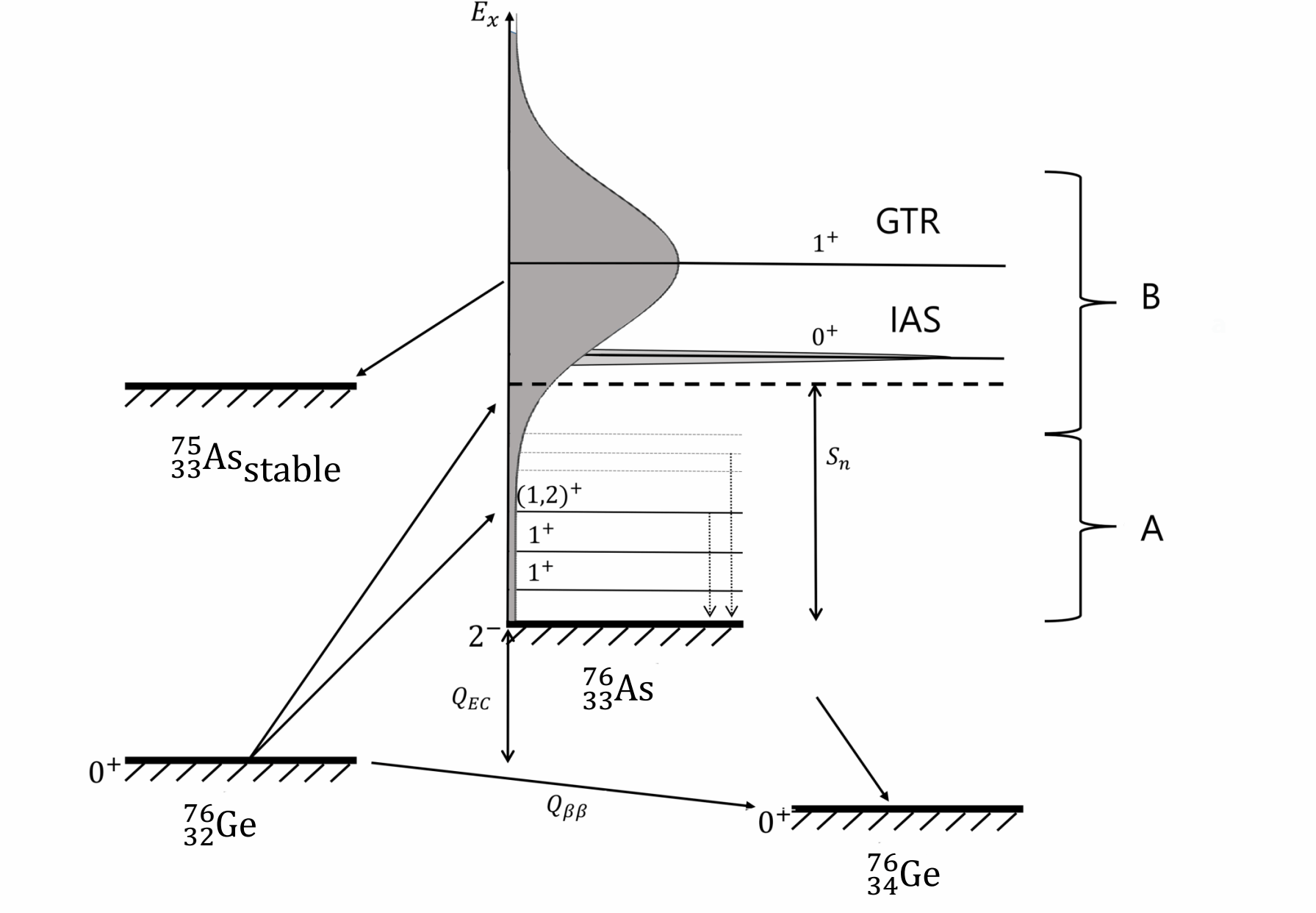}
	\caption{ Scheme of double beta decay of $\ce{^{76}Ge}$ through the intermediate $\ce{^{76}As}$ nucleus. The index A denotes the region of discrete excited levels; B is the energy region corresponding to the continuous excited states. The dashed line denotes the neutron separation energy $S_n$. Above are the analog IAS and Gamow-Teller - GTR resonances. }
	\label{Pic_1}
\end{figure}

One of the problems to be solved in modern and planned experiments on the search for double neutrinoless beta decay ($0\nu\beta\beta$ decay), is to determine the number of background events induced by solar neutrinos. Experiments like GERDA (or LEGEND) use detectors from ultra-pure crystalline $ \ce{^{76}Ge} $ as a target for studying the $ \ce{^{76}Ge} $ decay into $ \ce{^{76}Se} $. A statement of observation of the neutrinoless double beta decay of the $ \ce{^{76}Ge} $ nucleus could be made if the total energy of the emitted beta particles is 2039 keV  \cite{1}. The capture of solar neutrinos in the reaction $ \ce{^{76}Ge}( \nu, e)\ce{^{76}As} $  simulates this signal, thereby creating a practically unavoidable experimental background. In the first stage of the GERDA experiment  \cite{2}, neutrino-induced background events contributed insignificantly to the overall background level. For next-generation experiments (LEGEND)  \cite{3}, this question requires additional study, which raises additional interest in the topic of neutrino-nuclear reactions.
The influence of neutrino background events is shown in Fig. \ref{Pic_1}, where the excited states of the $ \ce{^{76}As}$ isotope, various parts of the excitation spectrum and isotopes formed as a result of neutrino capture and subsequent decays are schematically shown. As can be seen from Fig. \ref{Pic_1}, the $ \ce{^{76}Se} $ nucleus – the product of double beta decay – is also formed during neutrino capture by the initial $ \ce{^{76}Ge} $ nucleus, followed by the decay of the formed $ \ce{^{76}As} $ isobar into the final $ \ce{^{76}Se} $  \cite{4}. This process, induced by solar neutrinos, gives an appreciable number of background events, and we study it in this paper, including the resonance states of the continuous spectrum of charge-exchange excited states , which was not investigated previously. 

\section{Structure of    $\ce{^{76}As}$   excited states.}

The structure of the excitation levels of the isobaric nucleus in the charge-exchange reaction, for example, $(p, n)$, $(\ce{^{3}He}, t)$ or $(\nu_{e}, e^{–})$ can be determined using the charge-exchange strength function $S(E)$, where E is the excitation energy in the formed  isobaric nucleus. The region of discrete levels corresponds energies below 5 MeV. The region of continuous resonant states contains the wide giant Gamow-Teller resonance (GTR with $E_{GTR} \approx 11.3 MeV$  energy \cite{5}) and the analog resonance (isobaric analog state - IAS, narrow peak with energy $E_{IAS} = 8.308 MeV$ \cite{5}). The neutron separation energy is 
$S_n = 7.3285$ MeV  \cite{6}. For excited states with the energy of more than $S_n$, the  nucleus predominantly passes into stable isobar nuclei with A = 75, so here we focus on the $ \ce{^{76}As} $ underlying states. Despite the fact that $S_n$ is located below the position of the GTR peak ($S_n < E_{GTR}$), part of the resonant strength function $S(E)$ falls in the energy range of interest with $E < S_n$.

\section{ Cross section calculation for $(\nu_{e}, e^{–})$ reaction.}

The dependence of the total cross section $\sigma_{total}(E_{\nu})$ of the capture $(\nu_{e}, e^{–})$ reaction on the energy of the incident neutrino $ E_{\nu}$, taking into account the interaction with the continuous resonant states of the nucleus, is written as:
\begin{equation}\label{eq_1}
    \sigma_{total}(E_{\nu} ) = \sigma_{discr}(E_{\nu} )+ \sigma_{res}(E_{\nu} )   
\end{equation}

\begin{figure}[h!]
\center
	\includegraphics[width=1.0\linewidth]{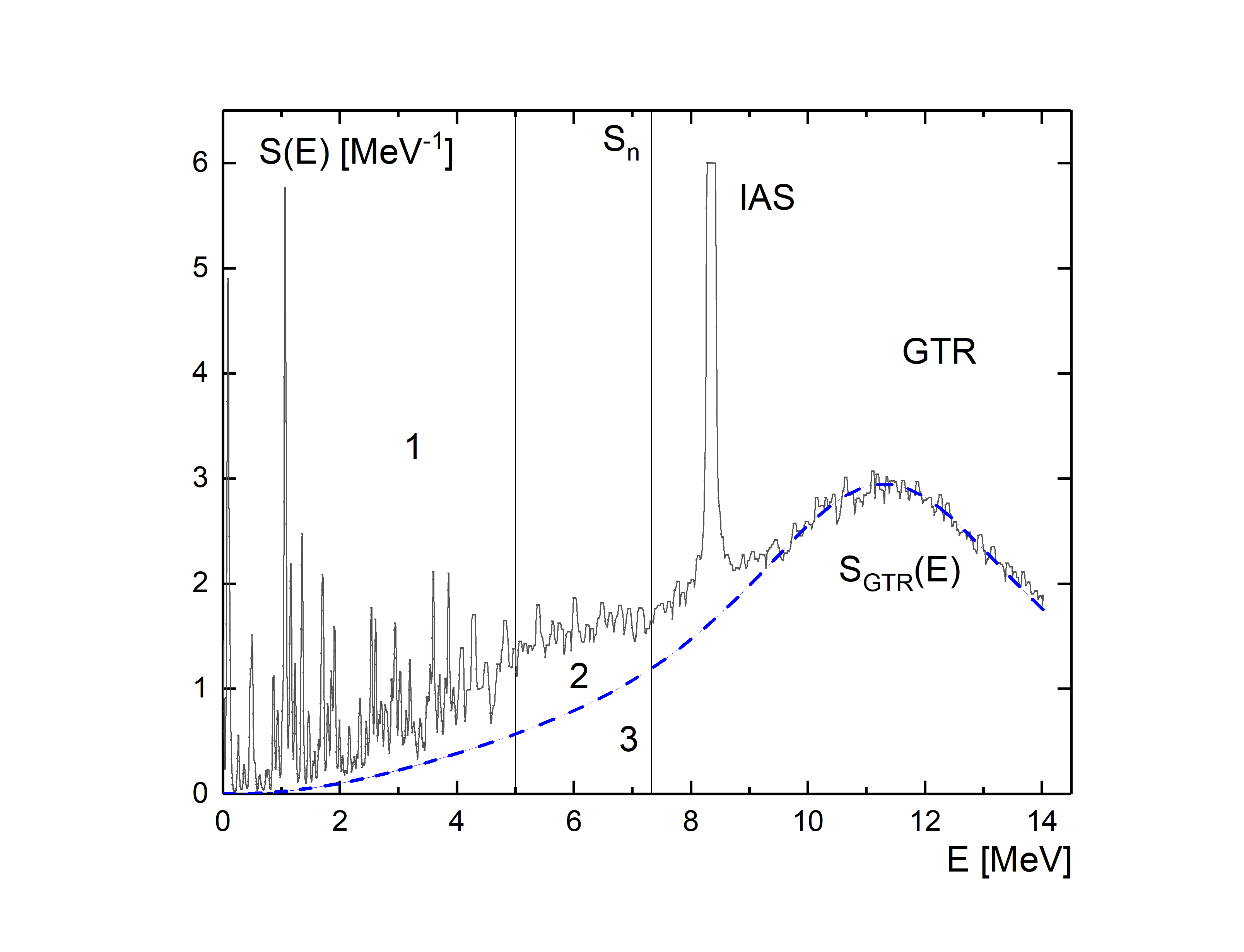}
	\caption{ 	Charge-exchange strength function  of the $ \ce{^{76}Ge}( \ce{^{3}He}, t)\ce{^{76}As} $ reaction $S(E)$  \cite{5}.
	The dashed line shows the Breit-Wigner approximation of the GTR. Vertical lines correspond to energies of 5 MeV and $S_n$. The digits denote areas: 1 - below 5 MeV, discrete levels; 2 - continuous excited states without GTR resonance; 3 - tail of the GTR resonance in the energy range from 5 MeV to $S_n$.}
	\label{Pic_2}
\end{figure}

Here $\sigma_{discr}(E_{\nu})$ is the discrete part determined by the tabular spectrum of excited states  \cite{1} with $E_{max} = 5$ MeV and it can be written as:
\begin{equation}
    \sigma_{discr}(E_{\nu}) = \frac{1}{\pi}  \sum\noindent_{k} G_F^2 \cos^2{\theta_{C}} \thinspace p_e E_e F(Z. E_e) [B(F)_k + (\frac{g_A}{g_V})^2 B(GT)_k  ]
\end{equation}
where $E_{e}, p_{e}$  - energy and momentum of the emitted electron, $ E_{e} - m_{e} c^2 = E_{\nu} -Q_{EC} - E > 0 $ , $F(Z, E_e)$ – the Fermi function  \cite{7}, $G_F$ – the Fermi constant, $\theta_C$ – the Cabibbo angle, $ B(F)_k$, $B(GT)_k$  – the Fermi and Gamow-Teller matrix elements  \cite{8}. The total cross section  is equal to zero for neutrino energies smaller than threshold energy $Q_{EC} = 921.51$ keV  \cite{6}. For  $E_{\nu}  > Q_{EC}$, the resonance cross section is defined as:
\begin{equation}
    \sigma_{res}(E_{\nu}) = \frac{1}{\pi} \int_{\epsilon_{min}}^{\epsilon_{max}} G_F^2 \cos^2{\theta_{C}} \thinspace p_e E_e F(Z, E_e) S(E) dE
\end{equation}
where the limits of integration $\epsilon_{min},\epsilon_{max}$  are given by the chosen energy range, where the contribution of the continuous charge-exchange strength function $S(E)$ is considered. In our estimation, $\epsilon_{min} = 5$ MeV, $\epsilon_{max} = S_n$.
Experimental data with resolution of 30 keV in the charge-exchange reactions on $ \ce{^{76}Ge} $ were obtained at the cyclotron of the University of Osaka in the $\ce{^{76}Ge}(\ce{^{3}He} ,t)\ce{^{76}As}$ reaction  \cite{5}. In particular, they recognized about 70 discrete excitation levels below the energy of 5 MeV (Fig. \ref{Pic_2}) and determined $B(GT)_k$. We use this data as an experimental strength function, both for the discrete and for the continuous part of the spectrum. Taking into account the sums for the Gamow-Teller transitions and  \cite{9}, we obtain:
\begin{equation}
    \sum\noindent_{i} M^2_i = \sum\noindent_{k} B(GT)_k + \int_{\Delta_{min}}^{\Delta_{max}} S(E)dE = 3\cdot (N-Z) = 36
\end{equation}
where $M_i^2$  is the square of the i-th matrix element, $\Delta_{min}  = 5$ MeV, $\Delta_{max}  = 28$ MeV –  the maximum experimentally known energy of the excitation spectrum for the intermediate nucleus. For the discrete part of the spectrum, our calculations reproduce the results  \cite{10} (tabular value  $\sum\noindent_{k} B(GT)_k = 1.6$ in  \cite{10}), and in the region above 5 MeV we distinguish two resonances of IAS (label 2 in Fig. \ref{Pic_2}) and GTR (label 3 in Fig. \ref{Pic_2}) and estimate their contribution to the strength function. For energies below $S_n$, the contribution of IAS turns out to be zero. The contribution of GTR is denoted by label 1 in Fig. \ref{Pic_2}. For the normalization of the continuous part of the excitation spectrum, the GT sum rule (4) is used.

The strength function $S(E)$  in the resonant energy region has the form:
\begin{equation}\label{qu5}
    S(E) = M_i^2 \cdot \frac{\Gamma_i}{(E - \omega_i)^2 + \Gamma_i^2}
\end{equation}
where the width $\Gamma_i$ is connected with the imaginary part of the self-energy operator  \cite{9}:
\begin{equation}\label{qu6}
    \Gamma(\epsilon) = -2 \operatorname{Im}(\epsilon + iI) = \alpha \epsilon^2 + \beta\epsilon^3 + \ldots
\end{equation}

In the calculations of $\Gamma(\epsilon)$ the parameter $\alpha$  takes into account the influence of three-quasiparticle configurations in the continuous spectrum. The numerical value of $\alpha \approx 0,018$ $MeV^{-1}$  is taken from  \cite{9}.
The influence of GTR in the experimental strength function of the $\ce{^{76}Ge}$ excited states $S(E)$ according to the paper \cite{5} we could determine as function  $S_{GRT}(E)$ like special case of (\ref{qu5}) with next parameters: $\omega_i = 11.3$ MeV, $\Gamma(\epsilon) = 3.3$ MeV (Fig. \ref{Pic_2}). Also we estimated $\beta = 6.9 \cdot 10^{-4} MeV^{-2}$  using the relation (\ref{qu6}) and data \cite{5}.

\section{Results}

\begin{figure}[h!]
\center
	\includegraphics[width=0.9\linewidth]{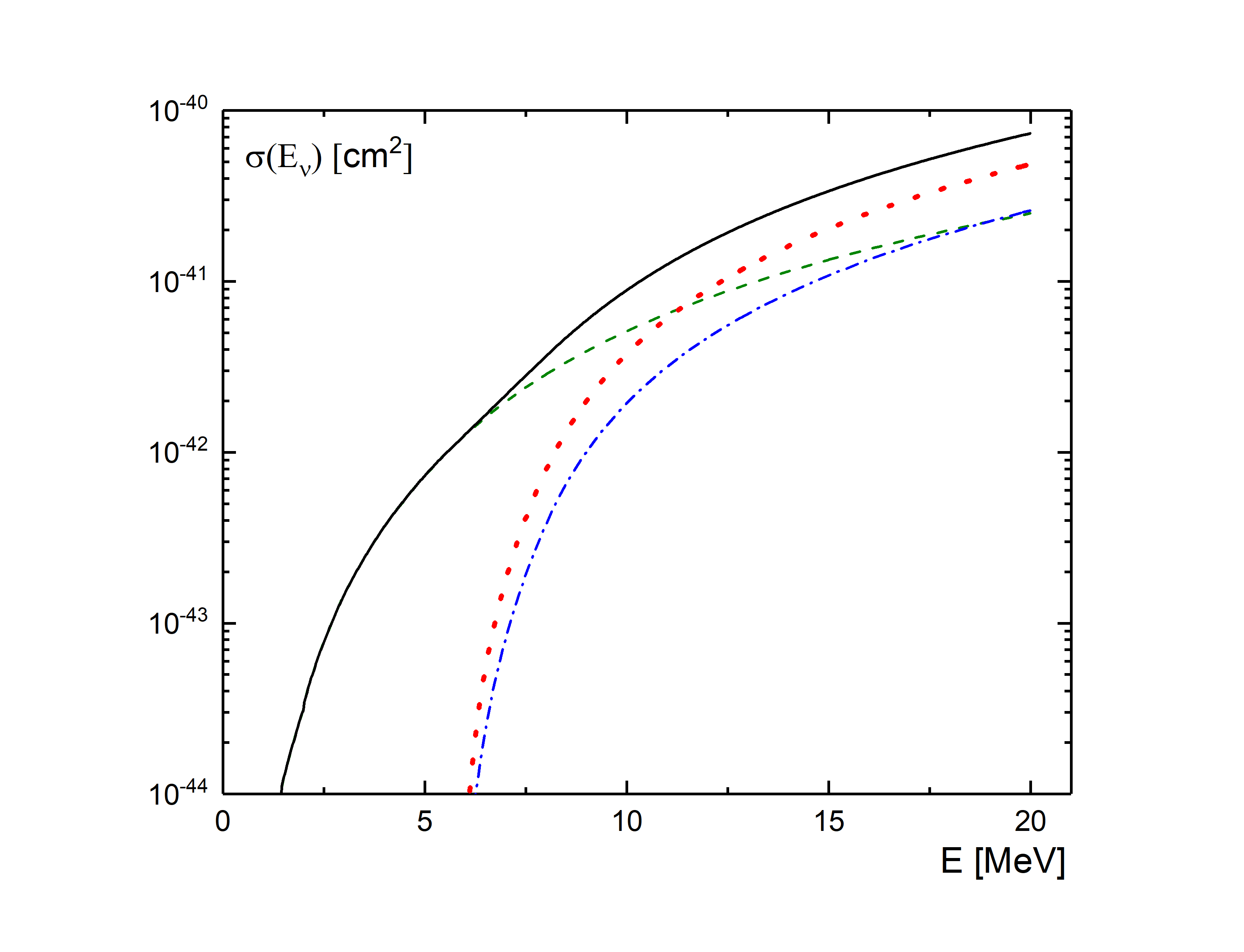}
	\caption{ 	The dependence of the neutrino capture cross sections on the energy of incident neutrinos: the red dotted line corresponds to the contribution from continuous excited states with energies from 5 MeV to $S_n$, the dashed green line from discrete states, the dash-dotted blue line is the contribution of GTR with energies from 5 MeV to $S_n$, the continuous line is the total cross section.    }
	\label{Pic_3a}
\end{figure}

\begin{figure}[h!]
\center
	\includegraphics[width=0.9\linewidth]{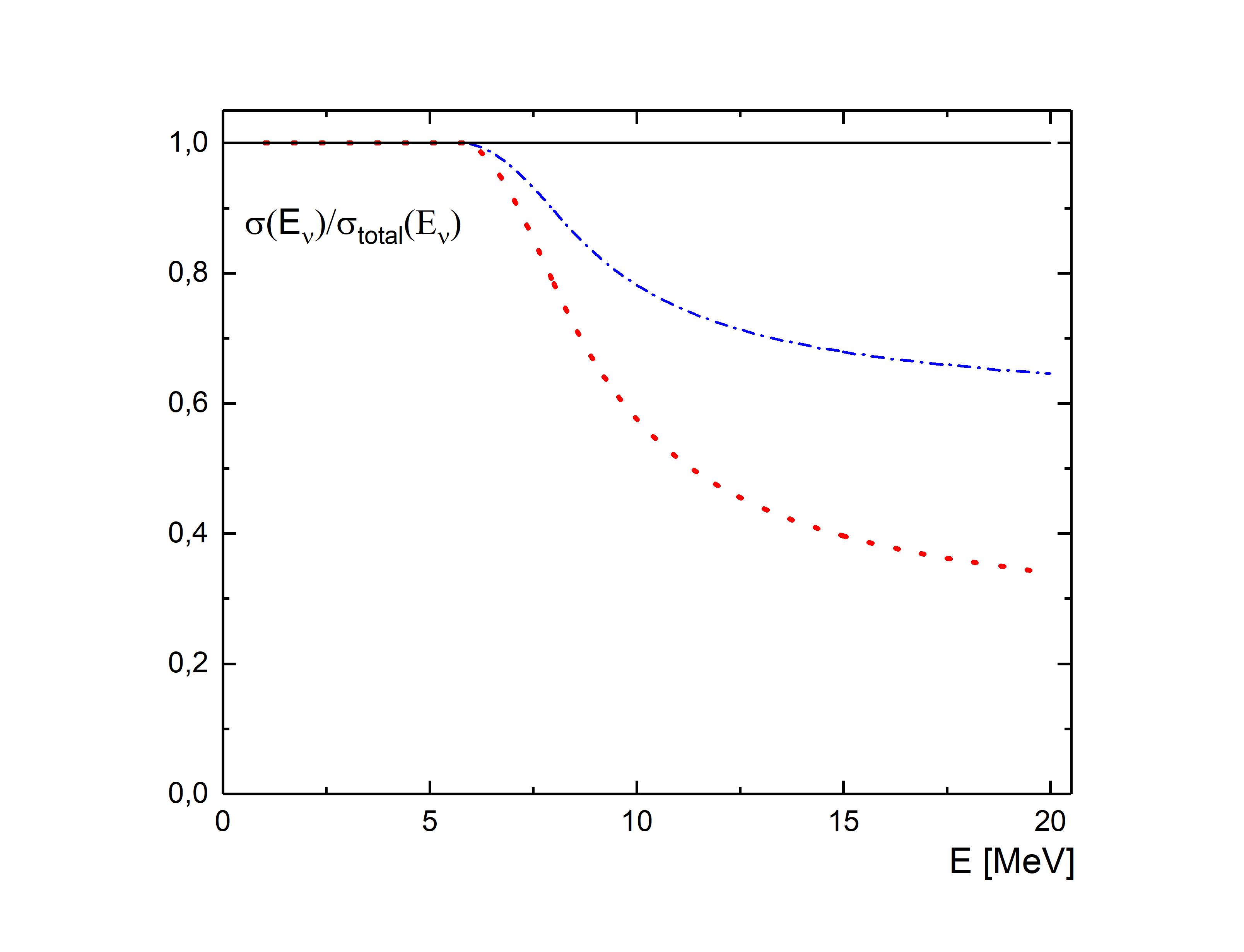}
	\caption{ 	Ratios of the cross section $\sigma(E_{\nu})$  for reaction $ \ce{^{76}Ge}( \nu, e)\ce{^{76}As} $ to the total cross section $\sigma_{total}(E_{\nu})$. 	$\sigma(E_{\nu})$  calculated in different assumptions of  the resonance structure influence: only discrete levels – dotted red line; discrete levels plus the GTR with energies from 5 MeV to $S_n$ – dash-dotted blue line.     }
	\label{Pic_3b}
\end{figure}

The dependence of the capture cross section of solar neutrinos on their energy is shown at Fig. \ref{Pic_3a}. The dashed line denotes cross sections for transitions to discrete levels below 5 MeV. The dotted line indicates cross sections for transitions to continuous states with energies from 5 MeV to $S_n  = 7.3285$ MeV. The dash-dotted line line indicates the contribution of only the GTR for the given energy range. A continuous line indicates the total cross section. The contribution of the continuous part of the spectrum becomes significant for incident neutrinos with energy above 6 MeV. Fig. \ref{Pic_3b} shows the fractions of the discrete part and the sum of the discrete part with the tail of GTR in the total cross section.

The neutrino capture rate is calculated like convolution of cross sections with flux densities of incident solar neutrinos taken using the model BS05(OP)  \cite{11}:
\begin{equation}
    R = \int_{0}^{E_{max}} \rho_{solar}(E_{\nu}) \sigma_{solar}(E_{\nu}) dE_{\nu}
\end{equation}
where $E_{max} = 18.79$ MeV is the maximum energy of the hep-neutrino formed in the reaction
$\ce{^{3}He} + p \rightarrow \ce{^{4}He} + e^{+} + \nu_e $. 

The results of calculation for each of the channels for the formation of solar neutrinos are summarized in the table \ref{Tab}. The main contribution to the final capture rate is provided by high-energy solar neutrinos obtained in the reaction: $\ce{^{8}B} \rightarrow \ce{^{8}Be^{*}} + e^{+} + \nu_e $. 
Neutrinos born in other channels, such as: $\ce{^{15}O} \rightarrow \ce{^{15}N} + e^{+} + \nu_e $,$\thinspace $ $ p + \thinspace  e^{-} + p  \rightarrow \ce{^{2}H} + \nu_e $ (pep), etc, do not make a significant contribution to the final capture rate. For transitions to discrete states below 5 MeV, the capture rate is 15.9 SNU, which is in complete agreement with  \cite{10} (15.6 SNU, where 1 SNU $= 10^{-36} \thinspace  (nucleon \cdot s)^{-1}$). Allowance for transitions to continuous states below the neutron separation energy increases the total capture rate by about 50$\%$, which is quite significant.

\section{Conclusions}

In this paper we calculated the capture rate of solar neutrinos by $\ce{^{76}Ge}$ nuclei. The neutrino capture cross sections $\sigma(E_{\nu})$ were determined by the method of nuclear strength functions $S(E)$ with allowance for the discrete and continuous states of the excited daughter isobaric nucleus $\ce{^{76}As}$. The effect of the resonance structure of the strength function $S(E)$ on the calculated cross section $\sigma(E_{\nu})$ was investigated. It is shown that only the giant Gamow-Teller resonance yields a contribution at least 20$\%$, and an even greater contribution is made by excited states located lower in the continuous part of the spectrum. The account of transitions to continuous states substantially increases the total capture rate by up to 50$\%$. These contributions should be considered in the calculation of background events in experiments on double beta decay of the GERDA type (LEGEND).

\begin{figure}[h!]
\center
	\includegraphics[width=0.9\linewidth]{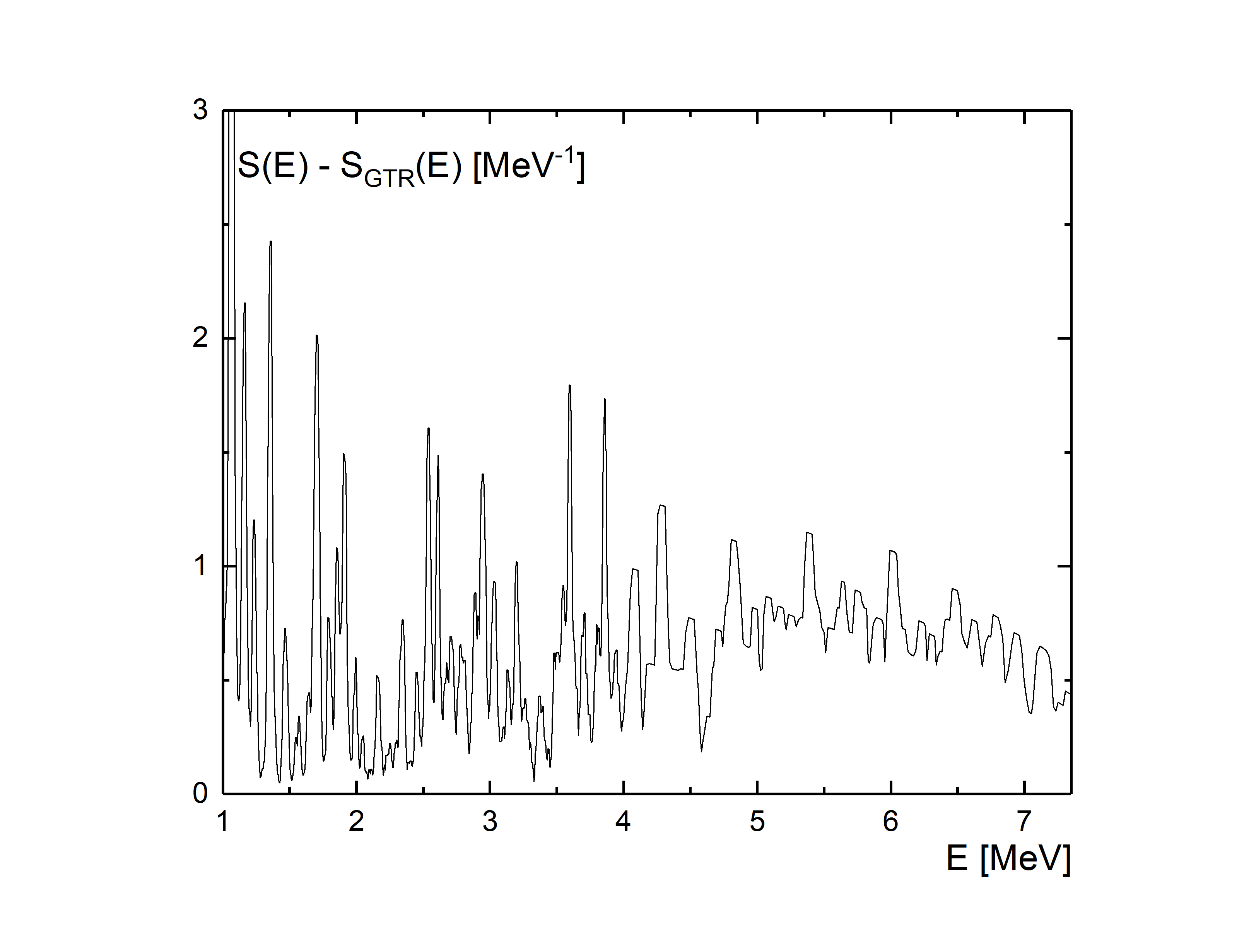}
	\caption{ 	Charge-exchange strength function of the $ \ce{^{76}Ge}( \ce{^{3}He}, t)\ce{^{76}As} $ reaction  $S(E)$ \cite{5} 
	minus contribution of the GTR.     }
	\label{difference}
\end{figure}

In the next stage of the research we are going to analyze influence of GT resonance on the area of discrete levels  and to determinate the pygmy resonance structure (see Fig. \ref{difference}). Also the methodology for calculating neutrino cross sections by taking into account the continuous states could be applied for the daughter nucleus. It is supposed to calculate the contribution of the secondary process $ \ce{^{76}As} \rightarrow \ce{^{76}Se} + e^{-} + \widetilde{\nu_e} $ to the background of solar neutrinos taking into account the design of the detecting elements.

\begin{table}[h!]
\centering
\resizebox{\textwidth}{!}{
\begin{tabular}{|l|l|l|l|l|l|l|l|}
\hline
  Capture rate  [SNU]                        & pep   & hep    & $\ce{^{13}N}$     & $\ce{^{17}F}$    & $\ce{^{15}O}$     & $\ce{^{7}B}$      & Total capture rate\\   \hline
$R_{total}$                    & 1.369 & 0.090  & 0.102 & 0.021 & 0.828 &  21.17 &  23.58            \\   \hline  
$R_{discr}$                    & 1.369 & 0.0451 & 0.102 & 0.102 & 0.102 & 0.102  & 15.9              \\   \hline
$R_{discr}$+$R_{GTR}$          & 1.369 & 0.070  & 0.102 & 0.021 & 0.828 & 17.46  & 19.85             \\   \hline
$(R_{total}-R_{GTR})/R_{total}$& 0$\%$ & 28$\%$ & 0$\%$ & 0$\%$ & 0$\%$ & 19$\%$ & 17$\%$            \\   \hline
\end{tabular}
}
\caption{The dependence of the capture rate of solar neutrinos on the model contribution of GTR through the neutrino generation channels in SNU: $R_{discr}$ units - only from discrete levels with energy below 5 MeV; $R_{discr}$ + $R_{GTR}$ - from discrete levels and the contribution of GTR in the energy range from 5 MeV to $S_n$; $R_{total}$ is the total capture rate with account of continuous states with energies from 5 MeV to $S_n$.}
\label{Tab}
\end{table}

The work was supported by the Russian Foundation for Basic Research, Grant No. 18-02-00670 and the state assignment of the Ministry of Education and Science of the Russian Federation 3.3008.2017 / PP.


\clearpage



\end{document}